Title: Singular Secular Kuznets-like Period Realized Amid Industrial Transformation in US FDA Medical Devices: A Perspective on Innovation from 1976 to 2020

Author: Iraj Daizadeh, PhD, Takeda Pharmaceuticals, 40 Landsdowne St. Cambridge, MA, 02139, iraj.daizadeh@takeda.com

Abstract:

Introduction: Since inception, the United States (US) Food and Drug Administration (FDA) has kept a robust record of regulated medical devices (MDs). Based on these data, can we gain insight into the innovation dynamics of the industry, including the potential for industrial transformation?

Areas Covered: Using Premarket Notifications (PMNs) and Approvals (PMAs) data, it is shown that from 1976 to 2020 the total composite (PMN+PMA) metric follows a single secular period: 20.5 years (applications – peak-to-peak: 1992-2012; trough: 2002) and 26.5 years (registrations – peak-to-peak: 1992-2019; trough: 2003), with a peak-to-trough relative percentage difference of 24% and 28%, respectively. Importantly, PMNs and PMAs independently present as an inverse structure.

Expert Opinion: The evidence suggests: MD innovation is driven by a singular secular Kutnets-like cyclic phenomenon (independent of economic crises) derived from a fundamental shift from simple (PMNs) to complex (PMAs) MDs. Portentously, while the COVID-19 crisis may not affect the overriding dynamic, the anticipated yet significant (~25%) MD innovation drop may be potentially attenuated with attentive measures by MD stakeholders. Limitations of this approach and further thoughts complete this perspective.

**Article Highlights**

- United States (US) Medical Device (MD) innovation may be tracked via novel metrics derived from publicly accessible data collected by the US Food and Drug Administration (FDA). The dynamics of these metrics seem to provide significant insight into innovation features of interest to researchers of the MD industry.

- The evolution of the FDA metrics – Premarket Notifications (PMNs; 510(k)) and Approvals (PMAs) – are individually temporally anticorrelated, suggesting an industrial structural shift in innovative activity (from PMNs to PMAs), due potentially (and speculatively) to more attractive economic rents irrespective of the higher relative regulatory burden.

- Together these metrics show a demonstrable singular secular Kuznets-like periodicity; the cyclicity of which is currently on a downward trend – reflecting a decrease in overall MD innovation – commencing prior to the COVID-19 crisis and anticipated to run for several years.

- All stakeholders involved in the MD development should be aware of such a potential phenomenon and strive to enable policies to buttress MD innovativeness for the sake of patients.

## 1. Introduction

### 1.1 Industrial Transformation as Technology Evolution

As a 'constructive-destructive' concept – as formulated by Schumpeter, his contemporaries, and thought progenies [1] –, when one thinks of an industrial transformation, typically one thinks about a change associated when a technology (or a suite of technologies) is temporally (and either incrementally or radically) usurped by another (or others) due to one or more endogenous (e.g., profit) or exogenous (e.g., environmental) driving forces [2]. These theories are explanatory in the sense that they presume a drive towards some sort of equilibrium [1], defined multi-factorially: size/breadth (micro-, meso-, macro-environment), kinetics (temporality), complexity, and cluster-ness (entanglement) [3].

In the simplest sense, consider a simple two-technology system: MD-1 and MD-2. The flow of MD-1 to MD-2 may be best described via the change in a value in an index (let's say number of units sold over time) from higher (lower) to lower (higher); that is, the industrial transformation realized when the utilization of MD-1 completely transfers to MD-2 (and vice versa) in the marketplace over time. While easy to conceptualize with 2 technologies (as a sinusoidal curve), it is much more challenging to consider a suite of technologies (yet alone thousands) comprising a broad industry or sector. In such a case, realizing such equilibria is challenging considering the various driving influences: size (state/national/international), rate (months/years/decades/centuries), complexity (e.g., multiple inputs/outputs), and type of technologies (which may or may not be overlapping, yet define the marketspace). Theoretically, it has been conceptualized that such complicated industrial transformations flow in concert, viewed coarsely as macro-cycles with periodicities that typically are not associated with any specific event (such as crisis) yet temporally integrable over many [1, 3, 4].

## 1.2 Economic Cycles

A broad assortment of economic variables – such as rates of population growth, emi/immigration, construction intensity, interest rates, foreign trade, industrial (e.g., coal and iron) production, climate, (and most recently US FDA approved medicines [5] and medical devices [6, 7]), and so on – cyclically ebb and flow over time and space [3]. Economists have attempted to define the various stages of such pulsating phenomena [1, 3-5, 8]: for example, Juglar defined this periodicity over three phases: prosperity, crisis, and subsequent liquidation, and suggested the length of the cycle with crisis/liquidation taking 1-2 years, followed by a 6-7-year phase of prosperity, with drivers to prosperity to crisis transition due to exuberance and thus over-speculation. Kitchin derived inventory cycles with wavelengths of 3.5 years. Kondratieff introduced the concept of the long-wave (50-60 year) cycles associated with inventiveness, while Kuznetz attributed medium-length cycles of 15-25 years to infrastructure concerns. Recently, some have argued the interconnectedness of these proposals (e.g., a sum of Kuznets swings equaling that of a Kondratieff cycle) [8]. Schumpeter argued (in part) via the 'creative destruction' hypothesis that a linear combination of industrial (technological) transformations drove such medium-to-long cycles using coarse macro-economic measures. While there are controversies yet to be resolved, and other models have been proposed, it is generally understood that cycles seem to exist albeit their precise mechanics have yet to be fully understood [1, 3-8].

## 1.3  United States (US) Food and Drug Administration (FDA, Agency) Regulated Medical Devices (MDs)

In the US, the FDA regulates MD (primarily under 21 CFR Parts 800-1299) [9]. Simplistically, the MD registration process is generally comprised of a submission of an application from the Sponsor followed by registration (upon successful review) by the Agency: categorized either as a Premarket Notification (PMN, 510k)[1] or approval (PMA)[2], the registration path depending on the risk profile of the intended use of the device in the intended population. A PMN requires that the applicant "demonstrate that the new device is 'substantially equivalent [SE]' to a predicate device in terms of intended use, technological characteristics, and performance testing, as needed[3]. PMNs also include those devices defined under De Novo Classification Scheme, wherein such an SE is not established as no predicate exists. A PMA, on the other hand, requires "valid scientific evidence demonstrating reasonable assurances of safety and effectiveness for the device's intended use [ibid]." To give the reader a taste of the diversity of MDs under each scheme: examples of the former include catheters, blood pressure cuffs, pregnancy test kits, syringes, etc..., while those of the latter include contact lenses, dermal fillers, and implantable systems [ibid].

---

[1] Note: MDs designated under the De Novo administrative cycle are integrated within PMNs.
[2] https://www.fda.gov/medical-devices

A given application (and its final regulatory disposition) may thus represent the culmination of substantial explicit (e.g., infrastructure/hard assets) and tacit (e.g., know-how/people) investments by the Sponsor(s) with the expectation of providing the market with a safe, effective/efficacious, quality, FDA-registered (cleared or approved) MD for an expected return. Thus, the gross number of applications or registrations over a given time may give significant insight into the evolution (pro/regress) of the regulated MD industry integrated over a diverse set of variables, e.g., economic, intellectual property (explicit/tacit), human resources, and/or policy matters [7].

**1.4 Investigating Industrial Transformations in US FDA Medical Devices**

In this work, we analyze the US MD landscape from the viewpoint of industrial transformation; effectively inquiring if the industry has or is undergoing a technology shift of some kind. For this analysis, the MD industry is considered holistically; the bases set defined by the gross number of FDA MD PMN and PMA applications or registrations. The benefit of such indices (beyond its conceptual utility) is the availability of the data, housed/maintained by a governmental organization that has been empowered to track and publicly share the regulatory status of specifically defined technologies. For the intended purpose here, the statistical analysis is simply to resolve a trendline; its existence would resolve qualitative temporal changes that may be indicative of a structural change in the industry.

---

[3] https://www.fda.gov/medical-devices/device-advice-comprehensive-regulatory-assistance/how-study-and-market-your-device#step2

Based on the results of this work, the author posits that such an industrial transformation has indeed occurred in the US FDA regulated MD industry; that is, the US FDA regulated MD industry is shifting from simple (PMNs) to complex (PMAs) devices. Further, as a collective (PMNs and PMAs) metric, a single Kuznetsian periodicity is observed independent of crises and suggests an innovation downturn is rapidly approaching. Additional results and interpretation complete this Expert Opinion. The author notes that all data and analytics routines are presented as an Appendix to this article to aid in reproducing, challenging, or extending this work.

## 2. Methods

### 2.1 Data and Data Analysis

Starting in May 1976, until the present, US FDA MD data is readily available and include: the number of applications registering for PMN and PMA status (collectively termed MD Applications), and the number attaining registration (collectively, termed MD Registrations). Independently, or collectively, these variables provide insight into the processes seeking and securing registration status for a MD in the US. For example, records comprising the number of MD Applications may reflect the innovativeness of developers, while that of MD Registrations may reflect the efficiency of the FDA in reviewing and approving the applications. The process for data collection has been provided in the Appendix and elsewhere [7].

The data analysis was comprised of constructing a yearly moving average to elucidate a trendline, which is subsequently qualitatively explored. The trendline was resolved using the Refined Moving Average Filter due to appropriateness of methodology, ease of access, and simplicity of execution [10].

The reader is directed to the Appendix for a copy of the data and R program code to exactly replicate, test, or extend the statistical analysis [10-15].

## 3. Results and Discussion

### 3.1 The Empirical Findings: PMNs and PMAs Inversely Related and Drive Periodicity

Figure 1 presents the monthly number of MD Applications (Figure 1a (PMN), c (PMA), and e (total)) and MD Registrations (Figures 1b (PMN), d (PMA), and f (total)). Importantly, qualitatively comparing the PMN curve (Figures 1a (Applications), 1b (Registrations)) with that of the PMA (1c (Applications), 1d (Registrations)) illustrates a dichotomous picture of PMN and PMA dynamics: effectively a reciprocal (inverse) relationship. The number of PMAs have been growing over the recent years; PMNs have been relatively stagnate (or overall decreasing) over the last decade. This is acutely observed in the trend-line depictions (left in red and right in green) estimated from a moving average calculation with a 12-month period. Superficially, it is also observed that there is a dip in 2020, the initial year of COVID-19, most visible in PMAs (but not in PMNs).

Of particular interest here is the clearly singular periodic (Bactrian) structure that manifests when both PMNs and PMAs are considered collectively (Figures 1e and f, and Figure 2). Figure 2 presents and defines a single period (that is, the date and amplitude of the two peaks and a trough) as:

- FDA MD Applications: Period: 20.5 years; Peak-to-Trough Relative Percent Difference: ~24%

  - Peak 1: Date: Apr 1992; Amplitude: 444

  - Peak 2: Date: Oct 2012; Amplitude: 442

- o  Trough: Date: Feb 2002; Amplitude: 339

- FDA MD Registrations: Period: 26.5 years; Peak-to-Trough Relative Percent Difference: ~28%

    - o  Peak 1: Date: Sept 1992; Amplitude: 433

    - o  Peak 2: Date: Mar 2019; Amplitude: 470

    - o  Trough: Date: Apr 2003; Amplitude: 340

## 4. Conclusion

From this analysis, the US MD innovative productivity was tracked via novel metrics derived from publicly accessible data collected by the US FDA. It is shown that the dynamics of these metrics provides significant insight into innovation features of the MD industry. The temporal evolution of the FDA metrics – PMNs and PMAs – are individually temporally anticorrelated, suggesting an industrial structural shift in innovative activity (from PMNs to PMAs), due potentially (and speculatively) to more attractive economic rents irrespective of the higher relative regulatory burden. Together these metrics show a demonstrable singular secular Kuznets-like periodicity; the cyclicity of which is currently on a downward trend – reflecting a decrease in overall MD innovative productivity – commencing prior to the COVID-19 crisis and anticipated to run for several years. The periodicity of the curve is roughly 20 years. All stakeholders involved in the MD innovative process should to be aware of such a potential phenomenon and strive for enabling policies to buttress MD discovery-to-delivery innovativeness for the sake of patients.

## 5. Expert Opinion

### 5.1 Findings 1 and 2: A single secular (approximately vicennial) Innovative (Kuznetsian) MD Period Due to a Shift from PMNs to PMAs

In the US, simplistically, MDs are regulated, and their regulatory status categorized based on complexity – effectively, the greater the complexity, the greater the degree of regulatory supervision. (Not unlike medicinal development), a complex (e.g., human-embedded) unprecedented MD, under the auspice of a PMA, may require a clinical trial to confirm safety, efficacy, and quality claims, while a simple precedented MD, under the auspice of the PMN (510k), generally do not. Axiomatically, the degree of MD complexity would increase proportionally to various innovation variables: access to advanced technologies and explicit (e.g., patents, publications) and tacit (e.g., human resource capabilities (know-how)) intellectual properties to construct and test the MD. Presumably, these infrastructure investments (which would grow proportional with MD complexity) would translate into economic rents (profits) to maintain and drive a firm's (or sectors) ongoing interest in product development. These infrastructure investments (particularly for the more complex production activities) take time to mature – potentially on the order of a vicennial as articulated by Kuznets [8].

Beginning in 1976 to present, the PMN/PMA regulatory construct has taken 40 years to evolve; maturing over a backdrop of significant economic events [8, 15], substantive policy changes [6, 7, 9, 16, 17], as well as major alterations in the tapestry of innovation levers (e.g., increases in governmental R&D expenditure [18], the volume of intellectual property [19, 20], and other sciento-economic metrics [21]). Economic recessions (1973-1975, 1980, and 1981-1982, and 2007) and shocks (notably, the 1987 Black Monday, the 2001 Dot-Com Crash, and the 2020 COVID-19 pandemic) have not been uncommon during this period [15], and MD policy changes rapidly promulgated [9, 16]: The 1976 amendments to the Federal Food, Drug, and Cosmetic (FD&C) Act, the 1990 Safe Medical Devices Act (SMDA), the 1992 Mammography Quality Standards Act (MQSA), 1997 FDA Modernization Act (FDAMA), the 2002 Medical Device User Fee and Modernization Act (MDUFMA) and its Reauthorization in 2007, 2012, and 2017, the 2016 21st Century Cures Act, the 2012 FDA Safety and Innovation Act (FDASIA), and a slew of responses to the COVID-19 crisis [17].

Gratefully, since its inception in the mid-1970s and irrespective of these external turbulences, the US FDA has been consistently supportive of maintaining and disseminating PMN/PMA records. These records describe respective innovations, regulatory categorization (PMA vs PMN), dates of application and registration, applicant names, and so forth. While there are limits to these data (such as limited to no cross-checking capability), relying on a key assumption they may be used as a 'system of record' to examine the industry allows for important investigations, including this one in which the PMN/PMA structure is examined.

As presented above, over these years, the gross number of PMNs have fallen while that of PMAs have concordantly risen, strongly suggesting an industry transformation / structural change defined by an industry producing more complex (PMAs) than simple (PMNs) MDs. Indeed, PMNs have effectively stagnated in growth for nearly 20 years (since ~2000), while PMAs enjoyed a (seemingly) exponential rise. As an aside, given that the regulatory requirement for PMNs include substantial equivalence of a precedence, thus the finding may be re reinterpreted as radical innovations supplanting those of incremental ones during the reporting period, reinforcing the concept that maturity in large infrastructural investment and perceived increase in economic rents driving the structural change in the industry.

Taken together as a composite index (PMN+PMA), a single secular period is observed for both applications and registrations, with peaks in the early 1990s and prior to 2020, with a trough around the turn of the century. The peak-to-peak (vicennial) distance falls well within a Kuznets swing – one in which industrial infrastructure cycle occurs [4, 6, 7]. There is always a temptation to ascribe inflection points with specific events. For example, the tough event may suggest an etiology to economic recovery post-dot-com crises. However, additional testing would need to be confirmed to test such a hypothesis.

In summary, using PMN/PMA as index, the evidence suggests: MD innovation is driven by a singular Kuznets-like secular cyclic phenomenon independent of economic crises derived from fundamental industrial shift from simple (PMNs) to complex (PMAs) MDs.

**5.2 Prognostic: A Steep (Natural) Decline in MD Productivity is Anticipated and Levers to Increase the Rate Innovation are Required Now**

For the PMN/PMA composite index, the peak-to-trough drop is roughly 25% in both applications and registrations and lasts roughly a decade. Given that the most recent peaks are roughly the late 2010's, assuming a continued business cycle, it would be anticipated that we are approaching a secular drop of roughly the same amount (precipitous) until the late 2020s. If realized, this may translate materially into lost quality of life and/or lives, given the importance of MDs in overall medical care – a dire consequence deserving focused attention by all stakeholders (across industry to government). While efforts are underway to consider 'silver linings' due to learnings from the most recent COVID-19 pandemic, it behooves all of us to think more actively about levers to optimize MD innovation. Importantly, this also means spending time / resources to further PMN innovation, which is stagnating.

In summary, portentously, while the COVID-19 crisis may not affect the overriding dynamic (as crises (theoretically) do not affect cyclicity), the anticipated yet significant (~25%) MD innovation drop (including PMN decay) may be attenuated with attentive policy measures.

**5.3 Study Limitations and Extensions**

While this study is preliminary and meant to drive a dialogue around PMN/PMA data and its utility in elucidating broad MD trends, including a reliance on a single source of metrics (which is challenging to cross-check for validation). Confidence exists based on the relatively low observed heteroscedasticity and from sporadic inspection of the records themselves. Regulatory evolution may result in re-classification or non-classification of MDs. While infrequent, it is possible that PMAs may become PMNs and PMNs may be further relegated overtime. Importantly, so-called software as a medical device (SaMD) is summarily exempt from the database due to its classification. Thus, it is possible that certain firms are moving toward producing SaMDs and therefore no longer contributing to PMNs. SaMDs may be supporting PMAs, however. Investigating – if possible – the innovation rate of SaMDs would be an future opportunity as this dimension of the industry evolves. Combination products (drug-device, device-device) may also be a motivating factor behind the rise of PMAs.

Additional extension of this work may include a formal inquiry into global MD trends would be warranted, given the firms' wishes to scale market share geographically and increase economic rents accordingly. Considering the impact of the ongoing COVID-19 pandemic (which is (by all recent signs) ebbing in the US) would be a critical factor to evaluate in the coming years [22].

In conclusion, a Kuznetsian middle-termed wave in the MD industry due to the inversely balanced dance between PMNs and PMAs is elucidated. This work also portends that we are in a relatively (decade-long) lasting recession in total MD development. Lastly, and importantly, if the forecast is correct, significant attention needs to be placed on PMN, as well as overall MD development.

**Declaration of Interest**

The author is an employee of Takeda Pharmacutics; however, this work was completed independently of his employment. The views expressed in this article may not represent those of his employer. See Appendix for all data and methods to replicate (test or extend) the results presented herein. To the author's knowledge, there are no relevant affiliations or financial involvement with any organization or entity with a financial interest in or financial conflict with the subject matter or materials discussed in the manuscript. This includes employment, consultancies, honoraria, stock ownership or options, expert testimony, grants, or patents received or pending, or royalties.

**Reviewer disclosures**

<<TBD>>

**Funding**

This paper was not funded.

**ORCID**

Iraj Daizadeh https://orcid.org/0000-0003-3648-023X

**References:**

Papers of special note have been highlighted as either of interest (•) or of considerable interest (••) to readers.

**Figure 1: The Observed Number (Black) of (a) FDA PMN Applications, (b) FDA PMN Registrations, (c) FDA PMA Applications, (d) FDA PMA Registrations, (e) FDA MD (PMN + PMA) Applications, and (f) FDA MD (PMN + PMA) Registrations with Respective Estimated Trendlines (Red on Left, and Green on Right)**

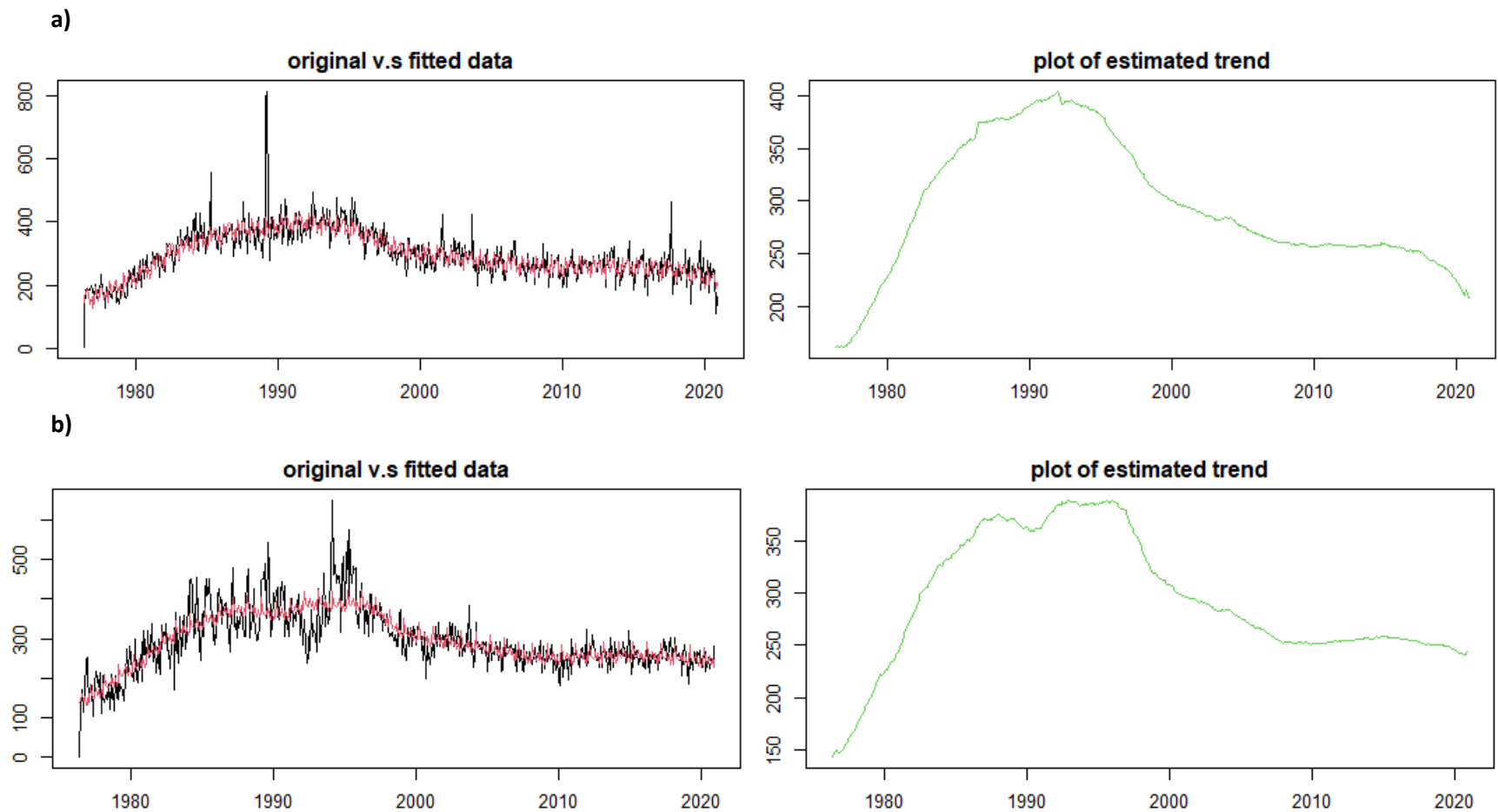

**c)**

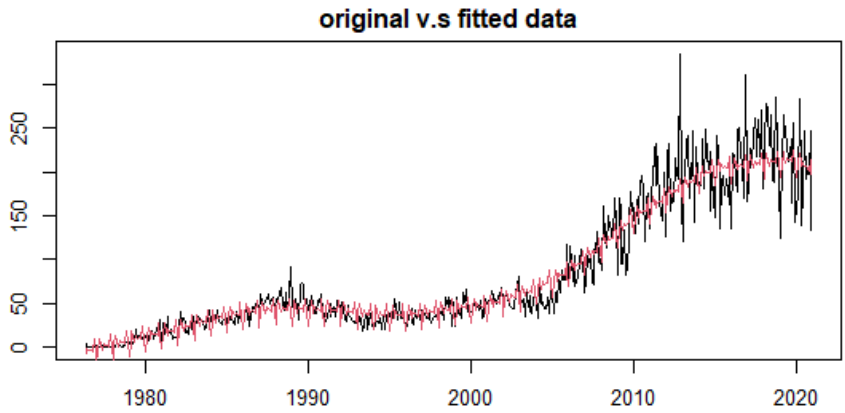
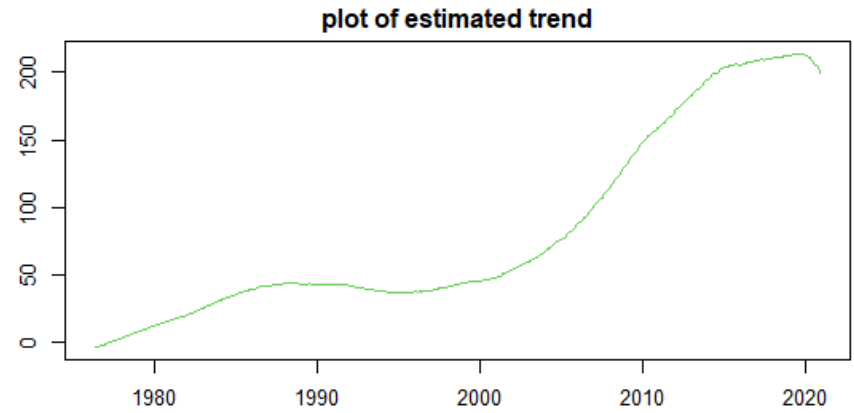

**d)**

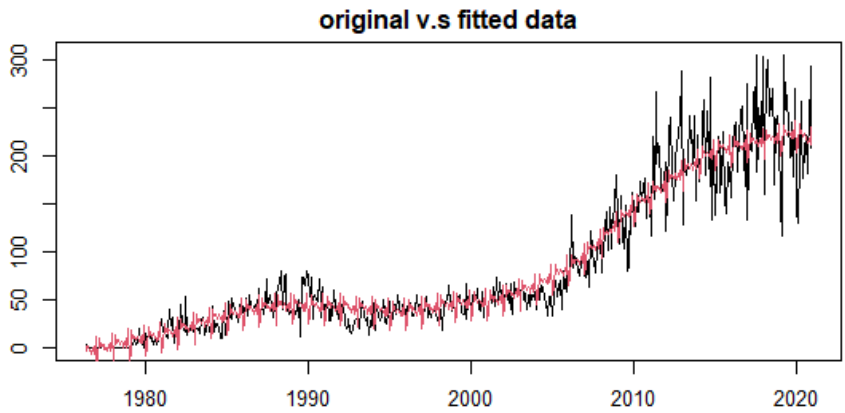
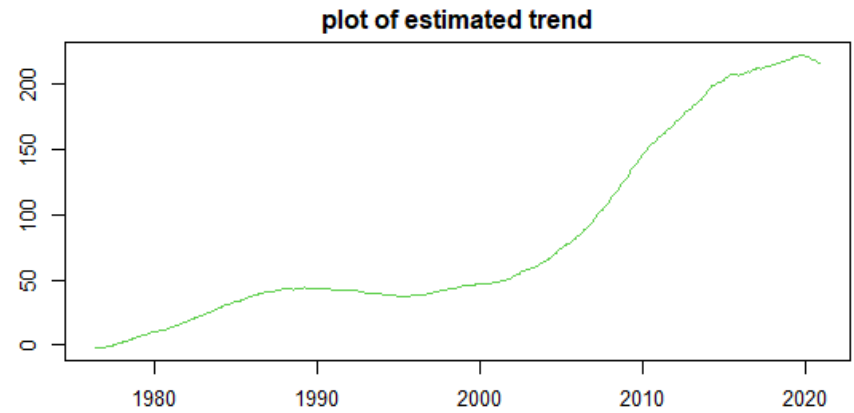

**e)**

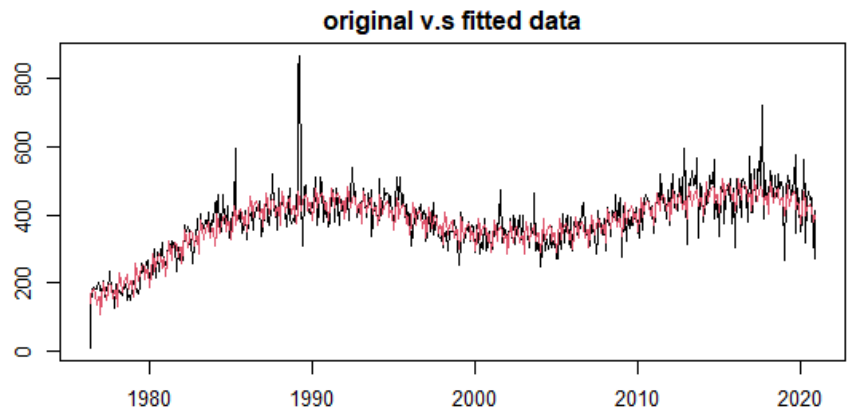
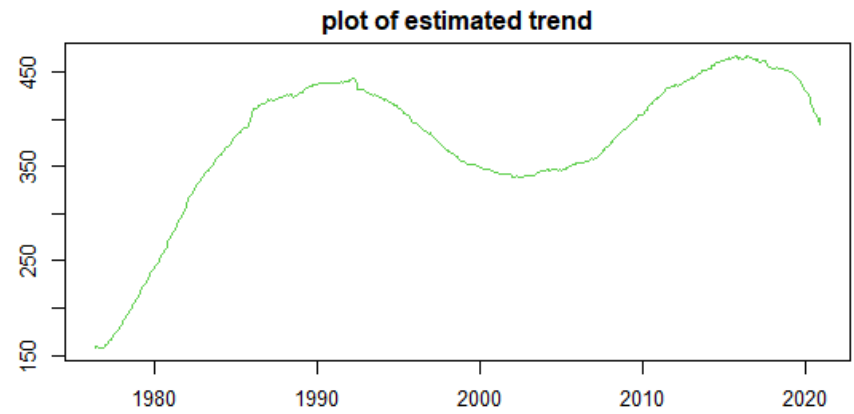

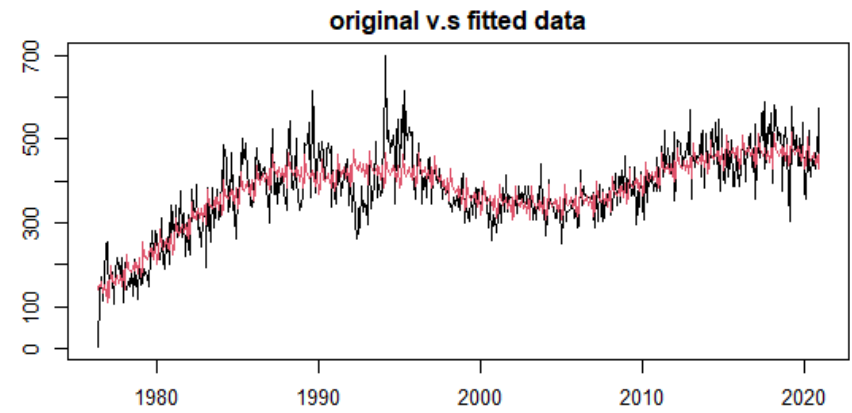
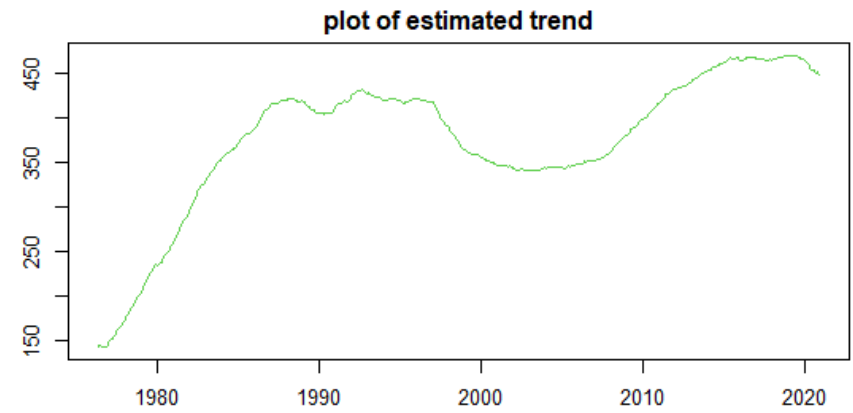

**Figure 2: Periods Resolved from Trends of the Number of FDA MD Applications (Red) and FDA MD Registrations (Blue): The Vertical (Horizontal) Lines Represent Date (Amplitude) of Local Maxima and Minima. Two Periods Identified are 20.5 Years (Red): Oct 2012 [442] - Apr 1992 [443], and 26.5 Years (Blue): Mar 2019 [470] - Sept 1992 [432].**

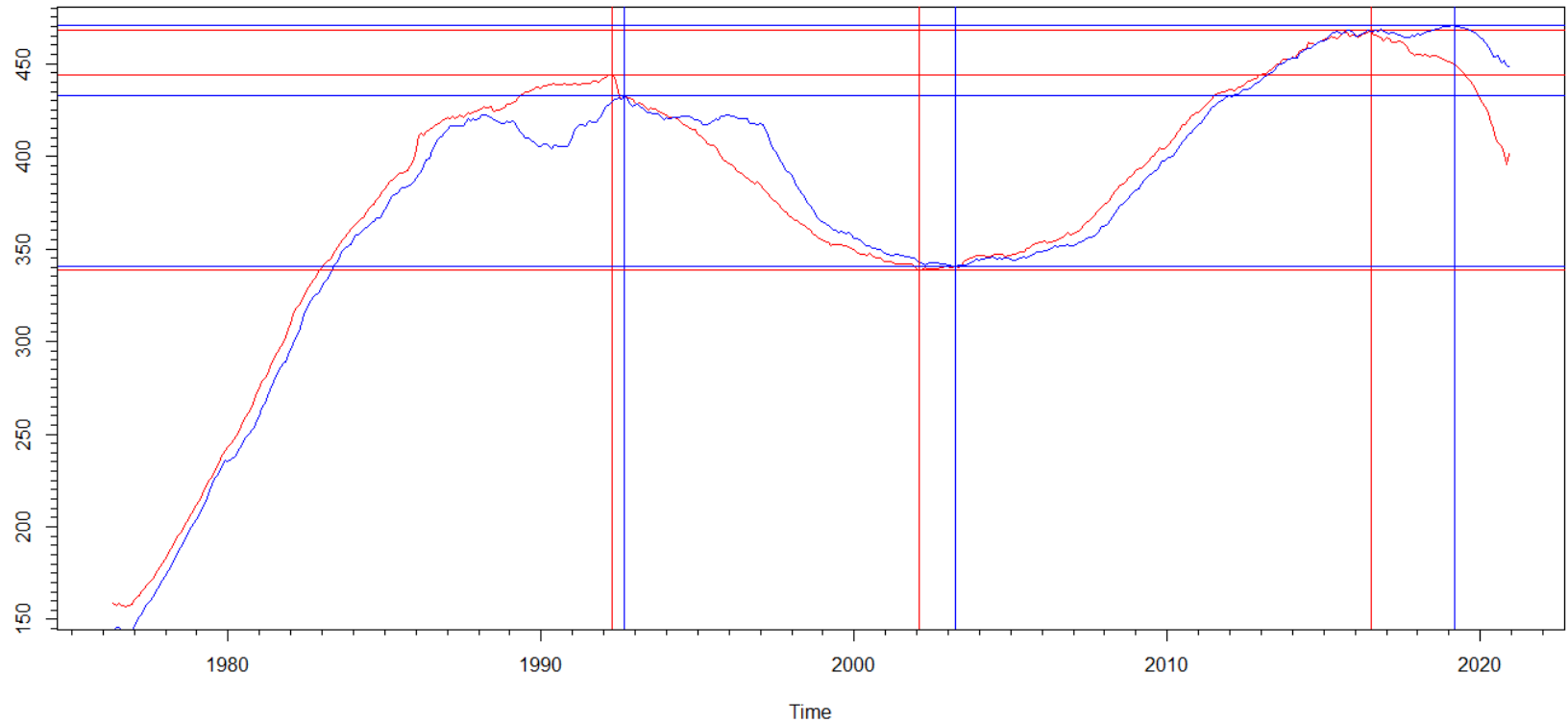

**Appendix 1: Details of Methodology and Statistical Analysis**

    **I.**      **Data Collection**

On 22-June-2021, Premarket Notifications (510(k); PMNs) and Premarket Approvals (PMA) data from the FDA websites: https://www.fda.gov/medical-devices/510k-clearances/downloadable-510k-files (1976-1980 (PMN7680.ZIP), 1981-1985 (PMN8185.ZIP), 1986-1990 (PMN8690.ZIP), 1991-1995 (PMN9195.ZIP), 1996-current (PMN96CUR.ZIP)) and https://www.fda.gov/medical-devices/device-approvals-denials-and-clearances/pma-approvals (PMA.zip under subsection: "PMA/PDP Files for Downloading"), respectively, were obtained. The data were read into EXCEL, and the monthly counts derived (by counting the number of PMNs or PMAs with the variables "DATERECEIVED" and "DECISIONDATE" within a given month). Comma delimited files (CSV) with the respective number of PMNs and PMAs were then imported into R for data analysis (see below).

    **II.**      **Statistical Analysis**

```
#Start
citation()
  R Core Team (2021). R: A language and environment for statistical computing. R Foundation for
  Statistical Computing, Vienna, Austria. URL https://www.R-project.org/.

version
platform      x86_64-w64-mingw32
arch          x86_64
os            mingw32
system        x86_64, mingw32
status
major         4
minor         1.1
year          2021
month         08
day           10
svn rev       80725
language      R
version.string R version 4.1.1 (2021-08-10)
nickname      Kick Things
```

```
#Read in file and process
Input<-read.csv("510k-PMA-R-TOTAL.csv", header=TRUE)
str(Input)
'data.frame':    536 obs. of  6 variables:
 $ X510kr    : int  3 128 184 189 183 184 200 184 190 164 ...
 $ X510kc    : int  0 0 130 170 114 168 250 253 199 157 ...
 $ PMAr      : int  4 0 0 0 0 0 0 3 1 0 ...
 $ PMAa      : int  4 2 0 1 0 1 0 0 0 0 ...
 $ X510krPMAr: int  7 128 184 189 183 184 200 187 191 164 ...
 $ X510kcPMAa: int  4 2 130 171 114 169 250 253 199 157 ...

Input
   X510kr X510kc PMAr PMAa X510krPMAr X510kcPMAa
1      3     0   4   4        7          4
2    128     0   0   2      128          2
3    184   130   0   0      184        130
4    189   170   0   1      189        171
5    183   114   0   0      183        114
6    184   168   0   1      184        169
7    200   250   0   0      200        250
8    184   253   3   0      187        253
9    190   199   1   0      191        199
10   164   157   0   0      164        157
11   205   180   0   0      205        180
12   159   183   0   0      159        183
13   186   104   3   2      189        106
14   184   193   0   0      184        193
15   169   180   1   0      170        180
16   233   217   0   0      233        217
17   175   198   0   0      175        198
18   176   178   1   1      177        179
19   128   215   0   0      128        215
20   196   110   0   1      196        111
21   145   205   0   0      145        205
```

| 22 | 160 | 155 | 0 | 0 | 160 | 155 |
| 23 | 179 | 140 | 0 | 1 | 179 | 141 |
| 24 | 169 | 158 | 4 | 0 | 173 | 158 |
| 25 | 160 | 147 | 2 | 0 | 162 | 147 |
| 26 | 193 | 185 | 2 | 0 | 195 | 185 |
| 27 | 193 | 123 | 1 | 0 | 194 | 123 |
| 28 | 171 | 211 | 3 | 0 | 174 | 211 |
| 29 | 147 | 151 | 0 | 0 | 147 | 151 |
| 30 | 158 | 175 | 3 | 1 | 161 | 176 |
| 31 | 142 | 118 | 6 | 0 | 148 | 118 |
| 32 | 169 | 207 | 8 | 1 | 177 | 208 |
| 33 | 203 | 150 | 4 | 2 | 207 | 152 |
| 34 | 180 | 223 | 7 | 2 | 187 | 225 |
| 35 | 189 | 147 | 5 | 7 | 194 | 154 |
| 36 | 168 | 186 | 4 | 3 | 172 | 189 |
| 37 | 160 | 175 | 7 | 8 | 167 | 183 |
| 38 | 173 | 171 | 20 | 4 | 193 | 175 |
| 39 | 245 | 142 | 10 | 5 | 255 | 147 |
| 40 | 241 | 203 | 13 | 21 | 254 | 224 |
| 41 | 195 | 270 | 13 | 6 | 208 | 276 |
| 42 | 222 | 274 | 16 | 8 | 238 | 282 |
| 43 | 244 | 214 | 11 | 7 | 255 | 221 |
| 44 | 262 | 279 | 6 | 1 | 268 | 280 |
| 45 | 189 | 257 | 13 | 16 | 202 | 273 |
| 46 | 238 | 214 | 9 | 3 | 247 | 217 |
| 47 | 216 | 191 | 7 | 20 | 223 | 211 |
| 48 | 280 | 233 | 18 | 13 | 298 | 246 |
| 49 | 240 | 300 | 19 | 11 | 259 | 311 |
| 50 | 194 | 199 | 19 | 11 | 213 | 210 |
| 51 | 282 | 187 | 9 | 4 | 291 | 191 |
| 52 | 221 | 246 | 13 | 10 | 234 | 256 |
| 53 | 263 | 226 | 23 | 4 | 286 | 230 |
| 54 | 295 | 292 | 21 | 7 | 316 | 299 |
| 55 | 268 | 191 | 22 | 9 | 290 | 200 |

| 56 | 238 | 321 | 31 | 22 | 269 | 343 |
| 57 | 245 | 283 | 6 | 30 | 251 | 313 |
| 58 | 243 | 234 | 27 | 17 | 270 | 251 |
| 59 | 305 | 321 | 16 | 19 | 321 | 340 |
| 60 | 284 | 254 | 18 | 12 | 302 | 266 |
| 61 | 285 | 273 | 36 | 11 | 321 | 284 |
| 62 | 281 | 279 | 14 | 17 | 295 | 296 |
| 63 | 301 | 336 | 17 | 40 | 318 | 376 |
| 64 | 292 | 243 | 21 | 23 | 313 | 266 |
| 65 | 227 | 291 | 9 | 32 | 236 | 323 |
| 66 | 290 | 214 | 12 | 6 | 302 | 220 |
| 67 | 262 | 260 | 11 | 22 | 273 | 282 |
| 68 | 247 | 299 | 14 | 19 | 261 | 318 |
| 69 | 247 | 236 | 22 | 7 | 269 | 243 |
| 70 | 242 | 211 | 23 | 14 | 265 | 225 |
| 71 | 329 | 334 | 40 | 45 | 369 | 379 |
| 72 | 321 | 303 | 28 | 26 | 349 | 329 |
| 73 | 330 | 309 | 33 | 19 | 363 | 328 |
| 74 | 328 | 328 | 29 | 17 | 357 | 345 |
| 75 | 295 | 337 | 17 | 54 | 312 | 391 |
| 76 | 309 | 271 | 19 | 13 | 328 | 284 |
| 77 | 243 | 270 | 16 | 18 | 259 | 288 |
| 78 | 284 | 223 | 29 | 19 | 313 | 242 |
| 79 | 301 | 290 | 20 | 22 | 321 | 312 |
| 80 | 352 | 296 | 36 | 20 | 388 | 316 |
| 81 | 307 | 327 | 12 | 14 | 319 | 341 |
| 82 | 276 | 172 | 23 | 21 | 299 | 193 |
| 83 | 375 | 364 | 29 | 21 | 404 | 385 |
| 84 | 320 | 287 | 36 | 20 | 356 | 307 |
| 85 | 325 | 238 | 34 | 15 | 359 | 253 |
| 86 | 355 | 354 | 29 | 31 | 384 | 385 |
| 87 | 337 | 255 | 29 | 32 | 366 | 287 |
| 88 | 382 | 299 | 21 | 26 | 403 | 325 |
| 89 | 384 | 331 | 25 | 13 | 409 | 344 |

| 90  | 327 | 360 | 23 | 18 | 350 | 378 |
| 91  | 348 | 275 | 36 | 31 | 384 | 306 |
| 92  | 338 | 302 | 43 | 17 | 381 | 319 |
| 93  | 400 | 408 | 29 | 17 | 429 | 425 |
| 94  | 408 | 342 | 30 | 23 | 438 | 365 |
| 95  | 357 | 442 | 31 | 44 | 388 | 486 |
| 96  | 427 | 451 | 34 | 17 | 461 | 468 |
| 97  | 317 | 426 | 33 | 26 | 350 | 452 |
| 98  | 307 | 301 | 43 | 25 | 350 | 326 |
| 99  | 429 | 319 | 30 | 22 | 459 | 341 |
| 100 | 356 | 456 | 27 | 11 | 383 | 467 |
| 101 | 341 | 320 | 26 | 10 | 367 | 330 |
| 102 | 341 | 402 | 27 | 12 | 368 | 414 |
| 103 | 395 | 284 | 31 | 39 | 426 | 323 |
| 104 | 300 | 246 | 32 | 14 | 332 | 260 |
| 105 | 345 | 363 | 27 | 17 | 372 | 380 |
| 106 | 367 | 307 | 40 | 48 | 407 | 355 |
| 107 | 374 | 433 | 31 | 34 | 405 | 467 |
| 108 | 554 | 409 | 40 | 30 | 594 | 439 |
| 109 | 369 | 449 | 31 | 52 | 400 | 501 |
| 110 | 362 | 411 | 30 | 46 | 392 | 457 |
| 111 | 362 | 449 | 26 | 41 | 388 | 490 |
| 112 | 370 | 377 | 41 | 36 | 411 | 413 |
| 113 | 318 | 366 | 28 | 27 | 346 | 393 |
| 114 | 314 | 295 | 44 | 46 | 358 | 341 |
| 115 | 319 | 338 | 36 | 36 | 355 | 374 |
| 116 | 382 | 286 | 31 | 51 | 413 | 337 |
| 117 | 301 | 317 | 28 | 41 | 329 | 358 |
| 118 | 355 | 403 | 29 | 33 | 384 | 436 |
| 119 | 355 | 426 | 50 | 52 | 405 | 478 |
| 120 | 372 | 391 | 37 | 34 | 409 | 425 |
| 121 | 333 | 397 | 24 | 43 | 357 | 440 |
| 122 | 355 | 339 | 44 | 56 | 399 | 395 |
| 123 | 362 | 372 | 31 | 46 | 393 | 418 |
| 124 | 373 | 387 | 47 | 39 | 420 | 426 |
| 125 | 382 | 390 | 44 | 44 | 426 | 434 |
| 126 | 391 | 352 | 49 | 46 | 440 | 398 |
| 127 | 290 | 298 | 46 | 40 | 336 | 338 |
| 128 | 354 | 275 | 47 | 62 | 401 | 337 |
| 129 | 318 | 267 | 35 | 32 | 353 | 299 |
| 130 | 349 | 401 | 60 | 36 | 409 | 437 |
| 131 | 407 | 478 | 54 | 46 | 461 | 524 |
| 132 | 341 | 334 | 48 | 35 | 389 | 369 |
| 133 | 318 | 324 | 52 | 46 | 370 | 370 |
| 134 | 365 | 355 | 57 | 72 | 422 | 427 |
| 135 | 342 | 301 | 55 | 43 | 397 | 344 |
| 136 | 462 | 345 | 60 | 49 | 522 | 394 |
| 137 | 365 | 336 | 49 | 56 | 414 | 392 |
| 138 | 374 | 377 | 54 | 46 | 428 | 423 |
| 139 | 301 | 336 | 57 | 32 | 358 | 368 |
| 140 | 430 | 306 | 47 | 53 | 477 | 359 |
| 141 | 341 | 307 | 55 | 24 | 396 | 331 |
| 142 | 337 | 391 | 51 | 53 | 388 | 444 |
| 143 | 412 | 442 | 56 | 63 | 468 | 505 |
| 144 | 386 | 475 | 60 | 68 | 446 | 543 |
| 145 | 341 | 303 | 68 | 80 | 409 | 383 |
| 146 | 315 | 338 | 50 | 40 | 365 | 378 |
| 147 | 388 | 344 | 42 | 75 | 430 | 419 |
| 148 | 415 | 426 | 63 | 77 | 478 | 503 |
| 149 | 345 | 353 | 46 | 44 | 391 | 397 |
| 150 | 333 | 329 | 42 | 41 | 375 | 370 |
| 151 | 329 | 294 | 61 | 36 | 390 | 330 |
| 152 | 287 | 297 | 91 | 39 | 378 | 336 |
| 153 | 403 | 336 | 47 | 34 | 450 | 370 |
| 154 | 423 | 382 | 46 | 49 | 469 | 431 |
| 155 | 784 | 443 | 42 | 46 | 826 | 489 |
| 156 | 813 | 452 | 56 | 35 | 869 | 487 |
| 157 | 424 | 451 | 46 | 40 | 470 | 491 |
| 158 | 277 | 490 | 32 | 50 | 309 | 540 |
| 159 | 415 | 350 | 69 | 12 | 484 | 362 |
| 160 | 415 | 544 | 73 | 73 | 488 | 617 |
| 161 | 343 | 470 | 72 | 73 | 415 | 543 |
| 162 | 370 | 472 | 56 | 68 | 426 | 540 |
| 163 | 361 | 306 | 53 | 65 | 414 | 371 |
| 164 | 347 | 334 | 55 | 80 | 402 | 414 |
| 165 | 386 | 416 | 43 | 75 | 429 | 491 |
| 166 | 394 | 395 | 49 | 54 | 443 | 449 |
| 167 | 454 | 420 | 59 | 73 | 513 | 493 |
| 168 | 329 | 431 | 41 | 30 | 370 | 461 |
| 169 | 387 | 358 | 53 | 52 | 440 | 410 |
| 170 | 355 | 347 | 38 | 56 | 393 | 403 |
| 171 | 473 | 416 | 38 | 51 | 511 | 467 |
| 172 | 434 | 443 | 54 | 48 | 488 | 491 |
| 173 | 327 | 396 | 51 | 72 | 378 | 468 |
| 174 | 357 | 433 | 60 | 49 | 417 | 482 |
| 175 | 382 | 305 | 57 | 32 | 439 | 337 |
| 176 | 372 | 361 | 44 | 63 | 416 | 424 |
| 177 | 331 | 336 | 35 | 29 | 366 | 365 |
| 178 | 349 | 307 | 29 | 38 | 378 | 345 |
| 179 | 429 | 377 | 33 | 54 | 462 | 431 |
| 180 | 428 | 402 | 47 | 28 | 475 | 430 |
| 181 | 365 | 329 | 49 | 40 | 414 | 369 |
| 182 | 345 | 337 | 36 | 28 | 381 | 365 |
| 183 | 434 | 359 | 43 | 36 | 477 | 395 |
| 184 | 378 | 345 | 46 | 60 | 424 | 405 |
| 185 | 335 | 386 | 35 | 42 | 370 | 428 |
| 186 | 391 | 411 | 53 | 39 | 444 | 450 |
| 187 | 373 | 309 | 50 | 31 | 423 | 340 |
| 188 | 363 | 337 | 41 | 56 | 404 | 393 |
| 189 | 359 | 365 | 46 | 44 | 405 | 409 |
| 190 | 384 | 285 | 52 | 43 | 436 | 328 |
| 191 | 407 | 349 | 42 | 47 | 449 | 396 |

| 192 | 350 | 284 | 33 | 25 | 383 | 309 |
| 193 | 422 | 240 | 30 | 20 | 452 | 260 |
| 194 | 491 | 257 | 48 | 22 | 539 | 279 |
| 195 | 496 | 304 | 39 | 30 | 535 | 334 |
| 196 | 424 | 271 | 35 | 18 | 459 | 289 |
| 197 | 436 | 373 | 34 | 15 | 470 | 388 |
| 198 | 364 | 307 | 29 | 21 | 393 | 328 |
| 199 | 391 | 318 | 28 | 28 | 419 | 346 |
| 200 | 405 | 312 | 32 | 30 | 437 | 342 |
| 201 | 391 | 268 | 23 | 28 | 414 | 296 |
| 202 | 383 | 296 | 22 | 37 | 405 | 333 |
| 203 | 444 | 426 | 35 | 20 | 479 | 446 |
| 204 | 409 | 291 | 34 | 41 | 443 | 332 |
| 205 | 375 | 315 | 18 | 48 | 393 | 363 |
| 206 | 442 | 352 | 24 | 35 | 466 | 387 |
| 207 | 390 | 465 | 39 | 29 | 429 | 494 |
| 208 | 437 | 358 | 34 | 31 | 471 | 389 |
| 209 | 318 | 330 | 21 | 32 | 339 | 362 |
| 210 | 353 | 378 | 28 | 13 | 381 | 391 |
| 211 | 396 | 336 | 38 | 28 | 434 | 364 |
| 212 | 400 | 352 | 26 | 24 | 426 | 376 |
| 213 | 355 | 486 | 22 | 47 | 377 | 533 |
| 214 | 361 | 602 | 25 | 18 | 386 | 620 |
| 215 | 475 | 650 | 31 | 50 | 506 | 700 |
| 216 | 369 | 461 | 30 | 40 | 399 | 501 |
| 217 | 376 | 486 | 27 | 35 | 403 | 521 |
| 218 | 444 | 439 | 33 | 29 | 477 | 468 |
| 219 | 437 | 458 | 20 | 32 | 457 | 490 |
| 220 | 429 | 457 | 36 | 51 | 465 | 508 |
| 221 | 435 | 422 | 22 | 41 | 457 | 463 |
| 222 | 373 | 317 | 20 | 27 | 393 | 344 |
| 223 | 419 | 475 | 37 | 25 | 456 | 500 |
| 224 | 368 | 509 | 32 | 37 | 400 | 546 |
| 225 | 329 | 384 | 39 | 49 | 368 | 433 |
| 226 | 396 | 492 | 31 | 24 | 427 | 516 |
| 227 | 475 | 546 | 34 | 36 | 509 | 582 |
| 228 | 406 | 406 | 55 | 45 | 461 | 451 |
| 229 | 393 | 573 | 54 | 41 | 447 | 614 |
| 230 | 463 | 484 | 50 | 35 | 513 | 519 |
| 231 | 394 | 440 | 31 | 56 | 425 | 496 |
| 232 | 410 | 482 | 51 | 46 | 461 | 528 |
| 233 | 363 | 475 | 33 | 33 | 396 | 508 |
| 234 | 347 | 479 | 41 | 39 | 388 | 518 |
| 235 | 399 | 377 | 33 | 40 | 432 | 417 |
| 236 | 355 | 349 | 37 | 52 | 392 | 401 |
| 237 | 382 | 328 | 20 | 30 | 402 | 358 |
| 238 | 283 | 400 | 26 | 39 | 309 | 439 |
| 239 | 314 | 440 | 33 | 48 | 347 | 488 |
| 240 | 353 | 374 | 41 | 47 | 394 | 421 |
| 241 | 374 | 368 | 44 | 43 | 418 | 411 |
| 242 | 348 | 311 | 30 | 34 | 378 | 345 |
| 243 | 363 | 373 | 26 | 34 | 389 | 407 |
| 244 | 376 | 370 | 42 | 39 | 418 | 409 |
| 245 | 397 | 306 | 33 | 55 | 430 | 361 |
| 246 | 348 | 410 | 43 | 39 | 391 | 449 |
| 247 | 381 | 411 | 32 | 45 | 413 | 456 |
| 248 | 357 | 316 | 42 | 31 | 399 | 347 |
| 249 | 295 | 324 | 27 | 20 | 322 | 344 |
| 250 | 301 | 376 | 33 | 36 | 334 | 412 |
| 251 | 368 | 385 | 38 | 41 | 406 | 426 |
| 252 | 320 | 358 | 33 | 53 | 353 | 411 |
| 253 | 357 | 363 | 33 | 32 | 390 | 395 |
| 254 | 396 | 370 | 40 | 40 | 436 | 410 |
| 255 | 307 | 389 | 36 | 39 | 343 | 428 |
| 256 | 335 | 374 | 28 | 32 | 363 | 406 |
| 257 | 366 | 318 | 34 | 38 | 400 | 356 |
| 258 | 307 | 356 | 36 | 43 | 343 | 399 |
| 259 | 293 | 334 | 37 | 33 | 330 | 367 |
| 260 | 303 | 362 | 34 | 32 | 337 | 394 |
| 261 | 276 | 320 | 29 | 25 | 305 | 345 |
| 262 | 291 | 325 | 33 | 27 | 324 | 352 |
| 263 | 302 | 305 | 41 | 40 | 343 | 345 |
| 264 | 309 | 309 | 41 | 18 | 350 | 327 |
| 265 | 265 | 316 | 32 | 45 | 297 | 361 |
| 266 | 295 | 296 | 57 | 44 | 352 | 340 |
| 267 | 304 | 310 | 49 | 55 | 353 | 365 |
| 268 | 292 | 281 | 57 | 56 | 349 | 337 |
| 269 | 325 | 298 | 47 | 66 | 372 | 364 |
| 270 | 329 | 282 | 53 | 47 | 382 | 329 |
| 271 | 331 | 359 | 24 | 38 | 355 | 397 |
| 272 | 273 | 374 | 52 | 44 | 325 | 418 |
| 273 | 229 | 259 | 25 | 40 | 254 | 299 |
| 274 | 267 | 310 | 56 | 37 | 323 | 347 |
| 275 | 349 | 292 | 58 | 48 | 407 | 340 |
| 276 | 348 | 262 | 38 | 52 | 386 | 314 |
| 277 | 281 | 247 | 38 | 55 | 319 | 302 |
| 278 | 308 | 327 | 33 | 38 | 341 | 365 |
| 279 | 299 | 282 | 63 | 35 | 362 | 317 |
| 280 | 327 | 328 | 52 | 35 | 379 | 363 |
| 281 | 256 | 318 | 66 | 67 | 322 | 385 |
| 282 | 292 | 280 | 63 | 30 | 355 | 310 |
| 283 | 325 | 329 | 43 | 49 | 368 | 378 |
| 284 | 325 | 320 | 38 | 61 | 363 | 381 |
| 285 | 249 | 283 | 40 | 37 | 289 | 320 |
| 286 | 327 | 329 | 42 | 52 | 369 | 381 |
| 287 | 278 | 331 | 51 | 59 | 329 | 390 |
| 288 | 260 | 315 | 46 | 52 | 306 | 367 |
| 289 | 240 | 319 | 47 | 50 | 287 | 369 |
| 290 | 268 | 294 | 48 | 62 | 316 | 356 |
| 291 | 257 | 247 | 42 | 40 | 299 | 287 |
| 292 | 336 | 317 | 54 | 39 | 390 | 356 |
| 293 | 263 | 198 | 34 | 58 | 297 | 256 |
| 294 | 275 | 287 | 36 | 32 | 311 | 319 |
| 295 | 252 | 271 | 59 | 41 | 311 | 312 |
| 296 | 296 | 262 | 53 | 48 | 349 | 310 |
| 297 | 242 | 236 | 50 | 40 | 292 | 276 |
| 298 | 246 | 269 | 53 | 44 | 299 | 313 |
| 299 | 312 | 289 | 53 | 55 | 365 | 344 |
| 300 | 290 | 253 | 65 | 47 | 355 | 300 |
| 301 | 314 | 321 | 60 | 64 | 374 | 385 |
| 302 | 299 | 329 | 50 | 53 | 349 | 382 |
| 303 | 338 | 281 | 56 | 44 | 394 | 325 |
| 304 | 422 | 341 | 49 | 74 | 471 | 415 |
| 305 | 260 | 280 | 59 | 47 | 319 | 327 |
| 306 | 303 | 298 | 58 | 68 | 361 | 366 |
| 307 | 286 | 308 | 68 | 54 | 354 | 362 |
| 308 | 283 | 289 | 56 | 66 | 339 | 355 |
| 309 | 265 | 305 | 44 | 46 | 309 | 351 |
| 310 | 265 | 308 | 53 | 45 | 318 | 353 |
| 311 | 326 | 291 | 62 | 67 | 388 | 358 |
| 312 | 274 | 271 | 56 | 52 | 330 | 323 |
| 313 | 349 | 340 | 50 | 47 | 399 | 387 |
| 314 | 266 | 290 | 46 | 67 | 312 | 357 |
| 315 | 341 | 324 | 47 | 45 | 388 | 369 |
| 316 | 300 | 336 | 44 | 35 | 344 | 371 |
| 317 | 325 | 282 | 44 | 68 | 369 | 350 |
| 318 | 336 | 331 | 64 | 57 | 400 | 388 |
| 319 | 258 | 289 | 69 | 55 | 327 | 344 |
| 320 | 317 | 318 | 80 | 71 | 397 | 389 |
| 321 | 261 | 300 | 43 | 49 | 304 | 349 |
| 322 | 279 | 255 | 48 | 52 | 327 | 307 |
| 323 | 326 | 300 | 52 | 69 | 378 | 369 |
| 324 | 289 | 255 | 56 | 51 | 345 | 306 |
| 325 | 270 | 280 | 36 | 57 | 306 | 337 |
| 326 | 308 | 266 | 65 | 40 | 373 | 306 |
| 327 | 306 | 303 | 47 | 50 | 353 | 353 |
| 328 | 282 | 322 | 59 | 61 | 341 | 383 |
| 329 | 424 | 291 | 40 | 44 | 464 | 335 |
| 330 | 258 | 384 | 39 | 56 | 297 | 440 |
| 331 | 245 | 271 | 56 | 53 | 301 | 324 |
| 332 | 283 | 289 | 61 | 58 | 344 | 347 |
| 333 | 197 | 229 | 52 | 40 | 249 | 269 |
| 334 | 267 | 250 | 33 | 66 | 300 | 316 |
| 335 | 286 | 340 | 66 | 61 | 352 | 401 |
| 336 | 277 | 283 | 73 | 50 | 350 | 333 |
| 337 | 247 | 291 | 49 | 65 | 296 | 356 |
| 338 | 276 | 280 | 51 | 63 | 327 | 343 |
| 339 | 253 | 285 | 45 | 53 | 298 | 338 |
| 340 | 277 | 264 | 45 | 46 | 322 | 310 |
| 341 | 312 | 293 | 53 | 39 | 365 | 332 |
| 342 | 247 | 278 | 57 | 33 | 304 | 311 |
| 343 | 274 | 288 | 39 | 54 | 313 | 342 |
| 344 | 269 | 292 | 70 | 54 | 339 | 346 |
| 345 | 206 | 216 | 63 | 34 | 269 | 250 |
| 346 | 235 | 264 | 38 | 56 | 273 | 320 |
| 347 | 285 | 284 | 63 | 71 | 348 | 355 |
| 348 | 243 | 246 | 57 | 56 | 300 | 302 |
| 349 | 268 | 264 | 51 | 54 | 319 | 318 |
| 350 | 324 | 264 | 77 | 63 | 401 | 327 |
| 351 | 257 | 260 | 62 | 71 | 319 | 331 |
| 352 | 291 | 293 | 89 | 40 | 380 | 333 |
| 353 | 322 | 273 | 77 | 81 | 399 | 354 |
| 354 | 246 | 245 | 77 | 65 | 323 | 310 |
| 355 | 214 | 257 | 92 | 59 | 306 | 316 |
| 356 | 276 | 290 | 118 | 93 | 394 | 383 |
| 357 | 218 | 218 | 76 | 69 | 294 | 287 |
| 358 | 261 | 222 | 115 | 71 | 376 | 293 |
| 359 | 287 | 288 | 101 | 139 | 388 | 427 |
| 360 | 268 | 256 | 76 | 84 | 344 | 340 |
| 361 | 263 | 298 | 104 | 90 | 367 | 388 |

| 362 | 309 | 278 | 70 | 95 | 379 | 373 |
|---|---|---|---|---|---|---|
| 363 | 284 | 277 | 75 | 93 | 359 | 370 |
| 364 | 340 | 280 | 81 | 81 | 421 | 361 |
| 365 | 338 | 284 | 97 | 75 | 435 | 359 |
| 366 | 247 | 271 | 111 | 91 | 358 | 362 |
| 367 | 257 | 262 | 86 | 96 | 343 | 358 |
| 368 | 249 | 283 | 103 | 90 | 352 | 373 |
| 369 | 266 | 246 | 62 | 76 | 328 | 322 |
| 370 | 232 | 217 | 75 | 73 | 307 | 290 |
| 371 | 273 | 283 | 104 | 91 | 377 | 374 |
| 372 | 252 | 242 | 98 | 63 | 350 | 305 |
| 373 | 244 | 271 | 107 | 121 | 351 | 392 |
| 374 | 212 | 219 | 73 | 103 | 285 | 322 |
| 375 | 264 | 216 | 72 | 86 | 336 | 302 |
| 376 | 299 | 286 | 111 | 93 | 410 | 379 |
| 377 | 260 | 233 | 108 | 79 | 368 | 312 |
| 378 | 244 | 266 | 132 | 103 | 376 | 369 |
| 379 | 247 | 241 | 97 | 112 | 344 | 353 |
| 380 | 287 | 267 | 101 | 116 | 388 | 383 |
| 381 | 205 | 238 | 88 | 79 | 293 | 317 |
| 382 | 239 | 244 | 88 | 104 | 327 | 348 |
| 383 | 260 | 234 | 162 | 95 | 422 | 329 |
| 384 | 282 | 257 | 123 | 126 | 405 | 383 |
| 385 | 226 | 278 | 113 | 141 | 339 | 419 |
| 386 | 260 | 242 | 150 | 101 | 410 | 343 |
| 387 | 255 | 263 | 139 | 147 | 394 | 410 |
| 388 | 287 | 244 | 133 | 104 | 420 | 348 |
| 389 | 331 | 266 | 129 | 134 | 460 | 400 |
| 390 | 244 | 295 | 125 | 130 | 369 | 425 |
| 391 | 252 | 212 | 170 | 150 | 422 | 362 |
| 392 | 303 | 280 | 145 | 180 | 448 | 460 |
| 393 | 192 | 225 | 82 | 132 | 274 | 357 |
| 394 | 247 | 265 | 170 | 108 | 417 | 373 |
| 395 | 278 | 275 | 159 | 158 | 437 | 433 |
| 396 | 294 | 242 | 95 | 157 | 389 | 399 |
| 397 | 226 | 249 | 95 | 117 | 321 | 366 |
| 398 | 277 | 266 | 133 | 103 | 410 | 369 |
| 399 | 279 | 285 | 82 | 148 | 361 | 433 |
| 400 | 282 | 213 | 92 | 81 | 374 | 294 |
| 401 | 297 | 254 | 122 | 85 | 419 | 339 |
| 402 | 227 | 254 | 142 | 122 | 369 | 376 |
| 403 | 220 | 219 | 177 | 119 | 397 | 338 |
| 404 | 266 | 261 | 156 | 162 | 422 | 423 |
| 405 | 195 | 192 | 159 | 149 | 354 | 341 |
| 406 | 203 | 183 | 131 | 126 | 334 | 309 |
| 407 | 252 | 275 | 168 | 155 | 420 | 430 |
| 408 | 245 | 249 | 141 | 141 | 386 | 390 |
| 409 | 196 | 203 | 162 | 145 | 358 | 348 |
| 410 | 259 | 243 | 185 | 173 | 444 | 416 |
| 411 | 262 | 247 | 196 | 153 | 458 | 400 |
| 412 | 268 | 249 | 181 | 166 | 449 | 415 |
| 413 | 314 | 197 | 121 | 177 | 435 | 374 |
| 414 | 239 | 260 | 171 | 135 | 410 | 395 |
| 415 | 251 | 210 | 176 | 167 | 427 | 377 |
| 416 | 264 | 299 | 160 | 158 | 424 | 457 |
| 417 | 211 | 260 | 135 | 157 | 346 | 417 |
| 418 | 221 | 257 | 181 | 117 | 402 | 374 |
| 419 | 278 | 265 | 185 | 219 | 463 | 484 |
| 420 | 236 | 254 | 228 | 155 | 464 | 409 |
| 421 | 239 | 253 | 185 | 267 | 424 | 520 |
| 422 | 290 | 241 | 232 | 191 | 522 | 432 |
| 423 | 266 | 226 | 204 | 213 | 470 | 439 |
| 424 | 267 | 271 | 169 | 208 | 436 | 479 |
| 425 | 279 | 278 | 154 | 166 | 433 | 444 |
| 426 | 243 | 253 | 145 | 163 | 388 | 416 |
| 427 | 275 | 246 | 200 | 175 | 475 | 421 |
| 428 | 272 | 322 | 159 | 193 | 431 | 515 |
| 429 | 245 | 230 | 126 | 122 | 371 | 352 |

| | | | | | |
|---|---|---|---|---|---|
| 430 | 248 | 240 | 218 | 152 | 466 | 392 |
| 431 | 289 | 274 | 232 | 225 | 521 | 499 |
| 432 | 267 | 256 | 155 | 239 | 422 | 495 |
| 433 | 249 | 308 | 170 | 183 | 419 | 491 |
| 434 | 244 | 283 | 166 | 154 | 410 | 437 |
| 435 | 296 | 269 | 210 | 182 | 506 | 451 |
| 436 | 288 | 257 | 216 | 200 | 504 | 457 |
| 437 | 273 | 247 | 197 | 208 | 470 | 455 |
| 438 | 236 | 264 | 192 | 224 | 428 | 488 |
| 439 | 259 | 228 | 335 | 234 | 594 | 462 |
| 440 | 306 | 281 | 162 | 288 | 468 | 569 |
| 441 | 192 | 226 | 121 | 129 | 313 | 355 |
| 442 | 223 | 244 | 184 | 177 | 407 | 421 |
| 443 | 295 | 283 | 196 | 188 | 491 | 471 |
| 444 | 254 | 251 | 235 | 180 | 489 | 431 |
| 445 | 289 | 266 | 241 | 192 | 530 | 458 |
| 446 | 316 | 213 | 170 | 241 | 486 | 454 |
| 447 | 303 | 249 | 211 | 215 | 514 | 464 |
| 448 | 264 | 303 | 206 | 211 | 470 | 514 |
| 449 | 322 | 220 | 247 | 191 | 569 | 411 |
| 450 | 187 | 279 | 143 | 242 | 330 | 521 |
| 451 | 266 | 228 | 228 | 154 | 494 | 382 |
| 452 | 286 | 292 | 196 | 221 | 482 | 513 |
| 453 | 198 | 260 | 183 | 173 | 381 | 433 |
| 454 | 231 | 250 | 185 | 188 | 416 | 438 |
| 455 | 225 | 263 | 210 | 207 | 435 | 470 |
| 456 | 249 | 278 | 238 | 235 | 487 | 513 |
| 457 | 253 | 267 | 182 | 258 | 435 | 525 |
| 458 | 267 | 261 | 248 | 180 | 515 | 441 |
| 459 | 272 | 292 | 219 | 247 | 491 | 539 |
| 460 | 288 | 269 | 203 | 205 | 491 | 474 |
| 461 | 338 | 238 | 224 | 169 | 562 | 407 |
| 462 | 234 | 265 | 156 | 282 | 390 | 547 |
| 463 | 240 | 242 | 199 | 134 | 439 | 376 |
| 464 | 275 | 319 | 169 | 204 | 444 | 523 |
| 465 | 187 | 273 | 149 | 138 | 336 | 411 |
| 466 | 227 | 210 | 241 | 183 | 468 | 393 |
| 467 | 278 | 243 | 224 | 220 | 502 | 463 |
| 468 | 236 | 281 | 193 | 211 | 429 | 492 |
| 469 | 239 | 248 | 136 | 161 | 375 | 409 |
| 470 | 266 | 268 | 188 | 176 | 454 | 444 |
| 471 | 292 | 271 | 195 | 225 | 487 | 496 |
| 472 | 284 | 233 | 174 | 155 | 458 | 388 |
| 473 | 314 | 246 | 193 | 140 | 507 | 386 |
| 474 | 194 | 264 | 181 | 194 | 375 | 458 |
| 475 | 242 | 230 | 163 | 178 | 405 | 408 |
| 476 | 270 | 257 | 210 | 188 | 480 | 445 |
| 477 | 169 | 231 | 135 | 156 | 304 | 387 |
| 478 | 284 | 225 | 222 | 209 | 506 | 434 |
| 479 | 245 | 281 | 219 | 229 | 464 | 510 |
| 480 | 257 | 242 | 191 | 235 | 448 | 477 |
| 481 | 216 | 214 | 178 | 184 | 394 | 398 |
| 482 | 250 | 261 | 246 | 191 | 496 | 452 |
| 483 | 257 | 239 | 250 | 216 | 507 | 455 |
| 484 | 272 | 254 | 222 | 246 | 494 | 500 |
| 485 | 252 | 260 | 210 | 252 | 462 | 512 |
| 486 | 211 | 237 | 170 | 202 | 381 | 439 |
| 487 | 261 | 254 | 311 | 168 | 572 | 422 |
| 488 | 288 | 259 | 185 | 275 | 473 | 534 |
| 489 | 241 | 222 | 167 | 133 | 408 | 355 |
| 490 | 227 | 248 | 181 | 193 | 408 | 441 |
| 491 | 305 | 274 | 227 | 217 | 532 | 491 |
| 492 | 231 | 249 | 209 | 205 | 440 | 454 |
| 493 | 270 | 298 | 242 | 261 | 512 | 559 |
| 494 | 297 | 292 | 261 | 272 | 558 | 564 |
| 495 | 263 | 247 | 237 | 183 | 500 | 430 |
| 496 | 258 | 282 | 218 | 305 | 476 | 587 |
| 497 | 461 | 247 | 259 | 196 | 720 | 443 |
| 498 | 174 | 285 | 217 | 241 | 391 | 526 |
| 499 | 230 | 303 | 271 | 209 | 501 | 512 |
| 500 | 256 | 260 | 181 | 303 | 437 | 563 |
| 501 | 221 | 234 | 223 | 160 | 444 | 394 |
| 502 | 205 | 244 | 236 | 208 | 441 | 452 |
| 503 | 252 | 301 | 278 | 280 | 530 | 581 |
| 504 | 237 | 268 | 275 | 299 | 512 | 567 |
| 505 | 216 | 255 | 230 | 242 | 446 | 497 |
| 506 | 251 | 264 | 266 | 251 | 517 | 515 |
| 507 | 261 | 244 | 192 | 269 | 453 | 513 |
| 508 | 271 | 252 | 189 | 212 | 460 | 464 |
| 509 | 312 | 207 | 209 | 170 | 521 | 377 |
| 510 | 192 | 257 | 286 | 234 | 478 | 491 |
| 511 | 269 | 285 | 228 | 242 | 497 | 527 |
| 512 | 288 | 256 | 184 | 223 | 472 | 479 |
| 513 | 142 | 229 | 124 | 147 | 266 | 376 |
| 514 | 258 | 186 | 231 | 116 | 489 | 302 |
| 515 | 249 | 272 | 247 | 304 | 496 | 576 |
| 516 | 251 | 223 | 266 | 218 | 517 | 441 |
| 517 | 251 | 249 | 243 | 277 | 494 | 526 |
| 518 | 226 | 221 | 234 | 252 | 460 | 473 |
| 519 | 261 | 267 | 212 | 240 | 473 | 507 |
| 520 | 280 | 237 | 221 | 199 | 501 | 436 |
| 521 | 338 | 238 | 238 | 200 | 576 | 438 |
| 522 | 184 | 267 | 164 | 235 | 348 | 502 |
| 523 | 224 | 259 | 256 | 179 | 480 | 438 |
| 524 | 286 | 270 | 143 | 270 | 429 | 540 |
| 525 | 194 | 232 | 157 | 144 | 351 | 376 |
| 526 | 208 | 228 | 176 | 130 | 384 | 358 |
| 527 | 279 | 275 | 284 | 202 | 563 | 477 |
| 528 | 237 | 265 | 191 | 257 | 428 | 522 |
| 529 | 176 | 216 | 140 | 226 | 316 | 442 |
| 530 | 270 | 232 | 177 | 177 | 447 | 409 |
| 531 | 224 | 253 | 246 | 206 | 470 | 459 |

```
532   250   223 193 221      443      444
533   248   249 202 181      450      430
534   108   246 198 258      306      504
535   163   229 247 208      410      437
536   135   279 134 293      269      572
```

```
library(tseries);
citation("tseries")
        Adrian Trapletti and Kurt Hornik (2020). tseries: Time Series
        Analysis and Computational Finance. R package version 0.10-48.
```

```
X510r<-ts(Input[1], start=c(1976,5), frequency=12)
X510c<-ts(Input[2], start=c(1976,5), frequency=12)
PMAr<-ts(Input[3], start=c(1976,5), frequency=12)
PMAa<-ts(Input[4], start=c(1976,5), frequency=12)
X510rPMAr<-ts(Input[5], start=c(1976,5), frequency=12)
X510cPMAa<-ts(Input[6], start=c(1976,5), frequency=12)
```

```
library(rmaf); citation("rmaf")
        Debin Qiu (2015). rmaf: Refined Moving Average Filter. R package version 3.0.1.
        https://CRAN.R-project.org/package=rmaf
```

```
library(pracma); citation("pracma")
        Hans W. Borchers (2021). pracma: Practical Numerical Math Functions. R
        package version 2.3.3. https://CRAN.R-project.org/package=pracma
```

```
MA510r<- ma.filter(X510r, seasonal = TRUE, period = 12, plot=TRUE)
MA510c<- ma.filter(X510c, seasonal = TRUE, period = 12, plot=TRUE)
MAPMAr<- ma.filter(PMAr, seasonal = TRUE, period = 12, plot=TRUE)
MAPMAa<- ma.filter(PMAa, seasonal = TRUE, period = 12, plot=TRUE)
MA510rPMAr<- ma.filter(X510rPMAr, seasonal = TRUE, period = 12, plot=TRUE)
MA510cPMAa<- ma.filter(X510cPMAa, seasonal = TRUE, period = 12, plot=TRUE)
```

```
library(zoo);citation("zoo")
        Achim Zeileis and Gabor Grothendieck (2005). zoo: S3 Infrastructure for Regular and Irregular
        Time Series. Journal of Statistical Software, 14(6), 1-27. doi:10.18637/jss.v014.i06

findpeaks(as.numeric(MA510rPMAr[,2], npeaks=2), threshold=2)
     [,1] [,2] [,3] [,4]
[1,] 443.6578 192  187  196 #1st peak
[2,] 467.0380 473  470  475
[3,] 467.8616 483  478  484 #2nd peak
[4,] 464.3752 489  488  491

plot(MA510rPMAr[,2], type="l");abline(h=t[192], v=as.yearmon(time(MA510rPMAr[,2])[192]), col="red"); abline(h=MA510rPMAr[,2][483],
v=as.yearmon(time(MA510rPMAr[,2])[483]), col="red")

findpeaks(as.numeric(MA510cPMAa[,2]), npeaks=200, threshold=0, minpeakheight=430)

      [,1] [,2] [,3] [,4]
 [1,] 431.2682 195  185  196
 [2,] 432.8777 197  196  200 #1st peak
 [3,] 432.5077 429  369  430
 [4,] 453.2280 454  430  455
 [5,] 467.3373 470  455  472
 [6,] 467.9159 474  472  478
 [7,] 468.0173 482  478  484
 [8,] 467.8377 485  484  486
 [9,] 468.6537 487  486  490
[10,] 466.7583 491  490  498
[11,] 465.6404 499  498  500
[12,] 465.8909 501  500  502
[13,] 467.4589 505  502  506
[14,] 468.5262 508  506  509
[15,] 470.3957 515  509  531 #2nd peak
[16,] 454.2619 532  531  533
[17,] 451.9979 534  533  536
```

```
plot(MA510cPMAa[,2], type="l");abline(h=t[197], v=as.yearmon(time(MA510cPMAa[,2])[197]), col="red"); abline(h=MA510cPMAa[,2][515],
v=as.yearmon(time(MA510cPMAa[,2])[515]), col="red")

library(Hmisc); citation("Hmisc")
        Frank E Harrell Jr (2021). Hmisc: Harrell Miscellaneous. R package
        version 4.6-0. https://CRAN.R-project.org/package=Hmisc

#Final Plot
plot(MA510rPMAr[,2], type="l", col="red", ylab="");abline(h=t[192], v=as.yearmon(time(MA510rPMAr[,2])[192]), col="red");
abline(h=MA510rPMAr[,2][483], v=as.yearmon(time(MA510rPMAr[,2])[483]), col="red")
lines(MA510cPMAa[,2], type="l", col="blue");abline(h=t[197], v=as.yearmon(time(MA510cPMAa[,2])[197]), col="blue");
abline(h=MA510cPMAa[,2][515], v=as.yearmon(time(MA510cPMAa[,2])[515]), col="blue")
minor.tick(nx=10, ny=10)

#Distance: 6 years
# 2nd peak to 1st peak: Oct 2012 [442.0573] - Apr 1992 [443.6578]
as.yearmon(time(MA510rPMAr[,2])[438]) - as.yearmon(time(MA510rPMAr[,2])[192])
20.5 #20 years and 5 months: Mar 2019 [470.3957] - Sept 1992 [432.8777]
as.yearmon(time(MA510cPMAa[,2])[515]) - as.yearmon(time(MA510cPMAa[,2])[197])
26.5 #26 years and 5 months

#END
```